\documentclass[journal=jctcce,manuscript=article]{achemso}
\setkeys{acs}{articletitle = true}
\setkeys{acs}{chaptertitle = true}
\setkeys{acs}{abbreviations = false}
\setkeys{acs}{maxauthors=150}
\setkeys{acs}{etalmode=truncate}
\usepackage[version=4]{mhchem}
\usepackage{xcolor}
\usepackage{mathtools,xparse}
\usepackage{placeins}
\usepackage{graphicx} 
\usepackage{amsmath}
\usepackage{amssymb}
\usepackage{subcaption}
\usepackage{float}
\usepackage{braket}
\usepackage{setspace}
\usepackage[colorlinks=true, linkcolor=black, citecolor=black, urlcolor=black]{hyperref}
\usepackage{xr}
\title{Quantum error mitigation using energy sampling and extrapolation enhanced Clifford data regression}
\author{Zhongqi Zhao}
\affiliation{Department of Chemistry, University of Copenhagen, DK-2100 Copenhagen \O, Denmark.}

\author{Erik Rosendahl Kjellgren}
\affiliation{Department of Physics, Chemistry and Pharmacy, University of Southern Denmark, Campusvej~55, DK--5230 Odense M, Denmark.}
\email{kjellgren@sdu.dk}

\author{Sonia Coriani}
\affiliation{Department of Chemistry, Technical University of Denmark, Kemitorvet Building 207, DK-2800 Kongens Lyngby, Denmark.}

\author{Jacob Kongsted}
\affiliation{Department of Physics, Chemistry and Pharmacy, University of Southern Denmark, Campusvej~55, DK--5230 Odense M, Denmark.}

\author{Stephan P. A. Sauer}
\affiliation{Department of Chemistry, University of Copenhagen, DK-2100 Copenhagen \O, Denmark.}

\author{Karl Michael Ziems}
\affiliation{School of Chemistry, University of Southampton, Highfield, Southampton SO17 1BJ, United Kingdom.}
\alsoaffiliation{Department of Chemistry, Technical University of Denmark, Kemitorvet Building 207, DK-2800 Kongens Lyngby, Denmark.}
\email{K.M.Ziems@soton.ac.uk}

\begin{document}
\maketitle
\pagebreak
\begin{abstract}
\label{sec:abstract}
Error mitigation is essential for the practical implementation of quantum algorithms on noisy intermediate-scale quantum (NISQ) devices. This work explores and extends Clifford Data Regression (CDR) to mitigate noise in quantum chemistry simulations using the Variational Quantum Eigensolver (VQE). 
Using the H$_4$ molecule with the tiled Unitary Product State (tUPS) ansatz, we perform noisy simulations with the \textit{ibm\_torino} noise model to investigate in detail the effect of various hyperparameters in CDR on the error mitigation quality. Building on these insights, two improvements to the CDR framework are proposed. The first, Energy Sampling (ES), improves performance by selecting only the lowest-energy training circuits for regression, thereby further biasing the sample energies toward the target state. The second, Non-Clifford Extrapolation (NCE), enhances the regression model by including the number of non-Clifford parameters as an additional input, enabling the model to learn how the noisy–ideal mapping evolves as the circuit approaches the optimal one. Our numerical results demonstrate that both strategies outperform the original CDR.
\end{abstract}
\section{Introduction}
\label{sec:intro}

We are currently in the noisy intermediate-scale quantum (NISQ) era, characterized by quantum devices with tens to hundreds of qubits. These systems already exhibit quantum computational capabilities that may outperform classical methods for certain specialized tasks~\cite{Preskill2018quantumcomputingin,Ladd2010,Harrow2017-qc,Kim2023}. However, the absence of full quantum error correction makes NISQ devices highly susceptible to noise and decoherence, which limits the accuracy and reliability of their computational outcomes~\cite{Preskill2018quantumcomputingin}. Achieving quantum utility---the point where quantum computations yield practically useful and accurate results---therefore remains a major challenge, particularly for applications such as quantum chemistry and materials simulation that demand high-precision estimations of observables~\cite{Harrow2017-qc}.
Fully fault-tolerant quantum computers (FTQCs) capable of implementing comprehensive error correction are still beyond reach, as they require a substantial overhead in qubits and circuit complexity~\cite{Preskill2018quantumcomputingin,PRXQuantum.2.040335}. As an alternative, quantum error mitigation (QEM) has emerged as a practical strategy to enhance computational accuracy without the need for additional qubits or complex encoding schemes~\cite{aharonov2025importanceerrormitigationquantum}. By reducing the impact of noise through algorithmic or post-processing techniques, QEM provides a resource-efficient pathway toward more reliable quantum computation within the constraints of current NISQ hardware~\cite{Temme_2017,PhysRevA.52.R2493,1996,Kitaev_2003,RevModPhys.95.045005,Guo2024}.

One widely used QEM is zero-noise extrapolation (ZNE), which mitigates errors by running the same circuit at different noise levels and extrapolating the results back to a zero-noise scenario~\cite{Temme_2017,9259940,PhysRevA.102.012426,10313813,Cai2021,PhysRevX.8.031027,PhysRevX.7.021050}. Probabilistic error cancellation (PEC)\cite{PhysRevX.8.031027,Temme_2017} is another potential method, that first learns a noise model and then uses quasi-probabilities to invert it,  thereby statistically canceling the effects of noise. While powerful, it requires significant sampling and computational overhead, making it less practical for very noisy or large circuits~\cite{vandenBerg2023,PhysRevA.104.052607,PRXQuantum.4.040329,PhysRevX.8.031027}. Various methods have been proposed to tackle this bottleneck, including combinations with device-characterization and suppression approaches, as well as tensor network error mitigation~\cite{aharonov2025reliablehighaccuracyerrormitigation,filippov2023scalabletensornetworkerrormitigation,filippov2024scalabilityquantumerrormitigation}. In addition, various read-out error mitigation (REM) techniques have been developed based on the original proposal by Bravyi et al~\cite{PhysRevA.103.042605}. For example, Twirled Readout Error eXtinction (TREX) mitigates readout errors by randomly applying $X$ gates before measurement and adjusting the measurement results accordingly~\cite{PhysRevA.105.032620}. Furthermore, REM has been extended to account for gate error effects in the Ansatz-based error mitigation method~\cite{D4SC05839A,graulund2025cost}. In addition to these techniques, other approaches such as purification methods, learning-based error mitigation, and hybrid strategies have been proposed, providing alternative ways to further reduce noise in quantum circuits.~\cite{PhysRevX.11.041036,cai2021resourceefficientpurificationbasedquantumerror,PRXQuantum.4.010303,PhysRevA.105.022427,D2SC06019A,PhysRevA.104.052607,Bultrini2023unifying,zhong2025combiningerrordetectionmitigation,filippov2023scalabletensornetworkerrormitigation,filippov2024scalabilityquantumerrormitigation}

A key challenge in error mitigation is scalability: it is difficult to ensure that these techniques remain effective and affordable as the number of qubits increases substantially. To solve this issue, a learning-based error mitigation scheme called Clifford Data Regression (CDR) has been proposed by Czarnik et al.~\cite{Czarnik2021errormitigation}. In learning-based error mitigation,~\cite{PRXQuantum.2.040330,RevModPhys.95.045005} the effect of noise is characterized by training on classically simulatable quantum circuits that resemble the target, non-simulatable circuits. The CDR method leverages the fact that circuits composed primarily of Clifford gates can be efficiently simulated classically~\cite{gottesman1998heisenbergrepresentationquantumcomputers}. Noiseless measurements are obtained from a classical computer, while corresponding noisy measurements are collected from a quantum device. A regression model is trained on these data pairs to predict the noise-free value of an observable from its noisy counterpart~\cite{Czarnik2021errormitigation}. Further developments based on CDR have focused on improving frugality through carefully chosen training data and symmetries\cite{Czarnik2025improvingefficiency}, on unifying CDR with noise-scaling approaches such as ZNE,\cite{lowe2021unified} and on exploiting empirically observed bias–dispersion correlations for self-calibrating mitigation.\cite{hosseinkhani2025noiserobustestimationquantumobservables}  Beyond CDR, other learning-based QEM methods train neural networks or other machine learning models to map noisy measurement statistics or expectation values to their ideal counterparts.\cite{liao2024machine,kim2020quantum,10.1063/5.0274910,bennewitz2022neural} 

In this work, we investigate the hyperparameters of the traditional CDR approach for molecular simulations and propose two enhancements: Energy Sampling (ES), which filters the training circuits by selecting the lowest-energy samples, and Non-Clifford Extrapolation (NCE), which incorporates additional circuit feature into the regression model. Specifically, we focus on device noise simulation of the ground state energy of the H$_4$ molecule using the tiled Unitary Product State (tUPS) ansatz~\cite{burton2024accurategateefficientquantumansatze,D4FD00064A} with different amount of layers and thus varying circuit depth and accuracy.

\section{Theory}
\label{sec:theory}

\subsection{Variational Quantum Eigensolver}
The variational quantum eigensolver (VQE) is a prominent hybrid quantum-classical algorithm to obtain the ground state energy and wave function of a molecular system. 
The latter is defined by the molecular electronic Hamiltonian, which is given by
\begin{equation}
\hat{H}_e = \sum_{pq}h_{pq}\hat{E}_{pq} + \frac{1}{2}\sum_{pqrs}g_{pqrs} \left( \hat{E}_{pq} \hat{E}_{rs} - \delta_{qr} \hat{E}_{ps} \right) ,
\end{equation}
where $h_{pq}$ and $g_{pqrs}$ are one- and two-electron integrals, the indices $p$, $q$, $r$, and, $s$ refers to spatial molecular orbitals, and $\hat{E}_{pq} = \hat{a}^{\dagger}_{p,\alpha} \hat{a}_{q,\alpha} + \hat{a}^{\dagger}_{p,\beta} \hat{a}_{q,\beta} $ is the singlet excitation operator, with $\hat{a}^{\dagger}_{p,\sigma}$ and $\hat{a}_{q,\sigma}$ being the creation and annihilation operators, respectively, for an electron of spin $\sigma$~\cite{doi:https://doi.org/10.1002/9781119019572.ch1}.
\par
To implement VQE, a parameterized wave function (Ansatz) must first be constructed. Its general form can be expressed as follows:
\begin{equation}
\ket{\Psi(\theta)} = \hat{U}(\theta)\ket{\Psi_0},
\end{equation}
where $\ket{\Psi_0}$ is an initial wave function (often Hartree-Fock, $\ket{\Psi_\mathrm{HF}}$) and $\hat{U}(\theta)$ is a parametrized unitary, whose parameters are optimized in the VQE routine to yield the energy minimum for a given Hamiltonian. In this work, we employ the Ansatz dubbed ``tiled Unitary Product State'' (tUPS)\cite{burton2024accurategateefficientquantumansatze} which is based on (i) fermionic operators, enabling tUPS to conserve particle number and spin symmetries, and (ii) layered Ansatz construction with repeated operators, allowing exact wave function representation in the limit of infinite layers~\cite{10.1063/1.5133059}. While the usage of fermionic and repeated operators is common in many methods like factorized UCC, k-UpCCGSD~\cite{Lee2019} and fermionic-ADAPT~\cite{Grimsley2019,PhysRevA.108.052422}, tUPS arranges the fermionic operator in tiled blocks resulting in lower depth circuits.
The energy minimization problem using VQE and a tUPS Ansatz is as follows:
\begin{equation}
E=\min_\theta\bra{\Psi_{\mathrm{HF}}}\hat{U}_{\mathrm{tUPS}}^\dagger(\theta)\hat{H}\hat{U}_{\mathrm{tUPS}}(\theta)\ket{\Psi_{\mathrm{HF}}}.
\end{equation}
where we seek the ground state energy of the system by minimizing the expectation value of the Hamiltonian with respect to a parameterized unitary transformation $\hat{U}_{\mathrm{tUPS}}(\theta)$ acting on the Hartree–Fock reference state $\ket{\Psi_{\mathrm{HF}}}$.

\subsection{Clifford Data Regression}

Clifford gates are elements of the Clifford group, consisting of unitary operators ${U}$ that map any Pauli operation to another Pauli operation under conjugation~\cite{Gottesman_1998}. Mathematically, the Clifford group can be expressed as:
\begin{equation}
    \mathcal{C}_n = \{U \mid UPU^\dagger \in \mathcal{P}_n,\, \forall P \in \mathcal{P}_n \},
\end{equation}
where $P \in \mathcal{P}_n$ is any Pauli operation, and $\mathcal{P}_n$ denotes the Pauli group acting on $n$ qubits. Quantum circuits consisting exclusively of Clifford gates are classically simulable in polynomial time, as stated by the Gottesman–Knill theorem~\cite{gottesman1998heisenbergrepresentationquantumcomputers}. As the classic simulation of non-Clifford gates scales exponentially, the larger the amount of non-Clifford operations in a circuit, the harder it becomes to simulate it classically. 

\begin{figure}[H]
  \centering
  \includegraphics[width=1\textwidth, clip, trim= 83 45 145 45]{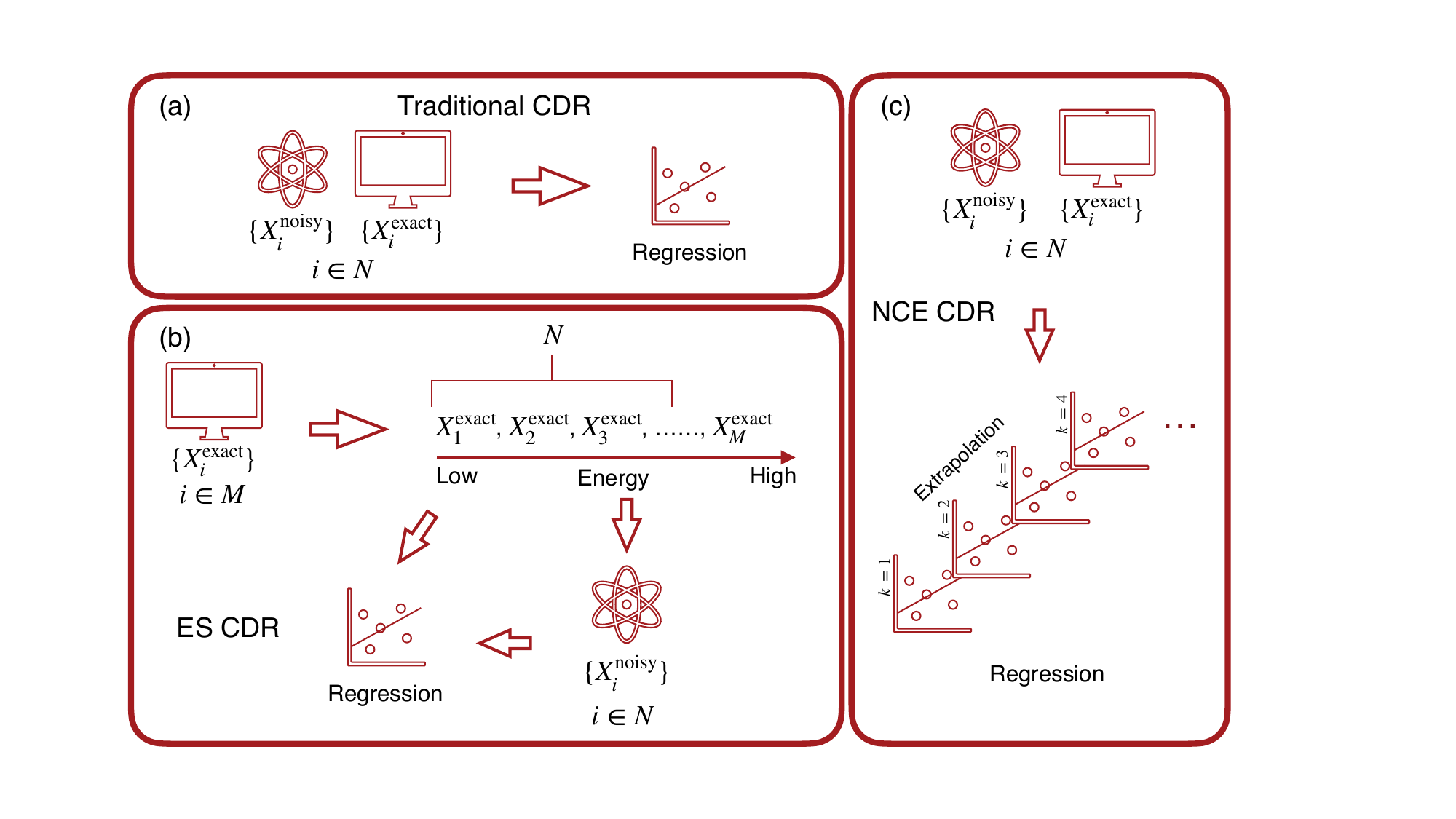}
  \caption{
    (a) Traditional CDR method: A set of (near-)Clifford circuits are executed on both a quantum device and a classical computer, and the resulting noisy and noise-free expectation values are used for regression.
    (b) Energy Sampling CDR: Noise-free expectation values from $M$ classically simulated (near-)Clifford circuits are obtained, and the $N<M$ lowest ones with their noisy counterparts are used for regression.
    (c) Non-Clifford Extrapolation CDR: A regression model incorporates the number of non-Clifford parameters $k$ as an additional feature in the training set, enabling the model to learn the dependence between noise-free and noisy expectation values in the low-$k$ regime and extrapolate to the target case $k=n$.
    }
  \label{fig:process}
\end{figure}

As shown in Fig.~\ref{fig:process}(a), the core idea behind CDR\cite{Czarnik2021errormitigation} is to train a regression model for hardware noise based on (near-)Clifford circuits and apply this to predict accurate expectation values for circuits containing a larger number of non-Clifford gates. The algorithm works as follows:
Suppose $\hat{X}$ is the operator of the observable of interest, and its exact expectation value in the given state $\ket{\Psi}$ is denoted as $X_\Psi^{\mathrm{exact}} = \bra{\Psi}\hat{X}\ket{\Psi}$. 
The value $X_\Psi^{\mathrm{noisy}}$ represents the corresponding expectation value obtained from a noisy device. Due to the presence of noise, $X_\Psi^{\mathrm{noisy}}$ typically deviates from $X_\Psi^{\mathrm{exact}}$. To mitigate this deviation, CDR trains a regression model based on near-Clifford versions of the original circuit, $\ket{\Psi}$.
To this end, a set of unique states $S_\Psi = \{ \ket{\Phi_i} \}$ is prepared to construct the training dataset $T$. To ensure that the expectation value of $X$ for each state can be efficiently computed on a classical computer, $\ket{\Phi_i}$ consists primarily of Clifford gates. Then, for each state $\ket{\Phi_i}$ in $S_\Psi$, $X_{\Phi_i}^{\mathrm{exact}}$ and $X_{\Phi_i}^{\mathrm{noisy}}$ are evaluated using a classical computer and a quantum device, respectively; the resulting pairs form the training dataset $T = \{ X_{\Phi_i}^{\mathrm{exact}}, X_{\Phi_i}^{\mathrm{noisy}} \}$. With the training dataset available, a regression model can be constructed, which learns how noise affects the expectation values by modeling the relationship between the noise-free and noisy results. Finally, one can input the noisy expectation value of the target state $X_\Psi^{\mathrm{noisy}}$ into the model to predict its counterpart mitigated expectation value $X_\Psi^{\mathrm{predicted}}$~\cite{Czarnik2021errormitigation}.

In chemically-inspired Ans\"atze for molecular simulation, rotation gates are the only type of parameterized non-Clifford gates. We refer to the phases of these non-Clifford gates as non-Clifford parameters. Given an Ansatz with $n$ non-Clifford gates, the circuits in $S_\Psi$ are prepared by converting $(n-k)$ non-Clifford gates to Clifford gates, while retaining a small portion of $k$ non-Clifford gates. This allows the generation of maximum $C(n,k)$ unique samples, with $C$ being the binomial coefficient function. In practice, a subset of $C(n,k)$ is chosen randomly. The simplest way to replace non-Clifford gates is to set their phases to $\theta = 0$. However, as mentioned in Ref.~\citenum{Czarnik2021errormitigation}, biasing the training dataset toward the target state can have a beneficial effect on the mitigation results. Thus, in this work we chose the Clifford gates whose phases are closest to the original phase, which we refer to as ``biasing'' in the following.

For the regression, the linear model was introduced in Ref.~\citenum{Czarnik2021errormitigation}. It is defined as:
\begin{equation}
f_{\mathrm{linear}}(X^{\mathrm{noisy}},a) = a_1 X^{\mathrm{noisy}} + a_2,
\end{equation}
where $a = \{a_1, a_2\}$ is a set of trainable parameters of the regression model. To fit the models, least squares regression was employed. The parameters $a$ are fitted by minimizing the cost function:
\begin{equation}
\label{eq:costfunction}
\mathcal{L}(a) = \sum_{\Phi_i \in S_\Psi} (X^{\mathrm{exact}}_{\Phi_i} - f(X^{\mathrm{noisy}}_{\Phi_i},a))^2.
\end{equation}
Once the parameters are obtained, the predicted expectation value $X_\Psi^\text{predicted}$ of the target state can be estimated by providing its noisy counterpart $X_\Psi^\text{noisy}$ as input,
\begin{equation}
X_\Psi^{\mathrm{predicted}} = f(X_\Psi^{\mathrm{noisy}},a).
\end{equation}

\subsection{Enhancement strategies}
The goal of this work is to enhance CDR with a variety of new strategies. These are introduced in the following.

\subsubsection{Quadratic regression}
First, we extend the regression function to using a quadratic model: \begin{equation}
f_{\mathrm{quadratic}}(X^{\mathrm{noisy}},a) = a_1 (X^{\mathrm{noisy}})^2 + a_2 X^{\mathrm{noisy}} + a_3,
\end{equation}
where $a = \{a_1, a_2, a_3\}$. The goal is to see whether the higher expressiveness is capable of capturing more complex nonlinear dependencies in the noise, which might allow for better fitting of the training data and reduce prediction error.

\subsubsection{Energy sampling}
\label{ES}
Second, we propose a pre-sampling strategy for the test set based on the energy of the classically simulated near-Clifford circuits (depicted in Fig.~\ref{fig:process}(b)). Meaning, we classically simulate $M$ samples and choose only the $N<M$ lowest-energy samples to build the training dataset $T^\mathrm{ES}$ (ES = energy sampling). As the energy is variationally bound, biasing the energies of training samples towards lower values might be able to better sample the space around the true ground-state energy. As only the $N$ lowest energy samples are put into the training set, the quantum cost does not increase. 

\subsubsection{Non-Clifford extrapolation}
\label{NCE}
Third, we use a regression model with a two-dimensional input, where the number of non-Clifford parameters $k$ in the circuits is added as an additional input feature (Non-Clifford extrapolation, NCE), illustrated in Fig.~\ref{fig:process}(c). The training set is then defined as $T^\mathrm{NCE} = \{ X_{\Phi_i}^{\mathrm{exact}}, X_{\Phi_i}^{\mathrm{noisy}}, k_i \}$. This is to allow for the model to capture how the relationship between the noise-free and noisy values varies in the low-$k$ regime, and then extrapolate to the case $k = n$, which corresponds to the circuit of the target state and yields the desired expectation value. The model is defined as 
\begin{equation}
\label{eq:kansatz}
f_{\mathrm{NCE}}(X^{\mathrm{noisy}},k,a) = a_1 (X^{\mathrm{noisy}})^2 + a_2 k^2 + a_3 k X^{\mathrm{noisy}} + a_4 X^{\mathrm{noisy}} + a_5 k + a_6,
\end{equation}
where $a = \{a_1, a_2, a_3, a_4, a_5, a_6\}$. The model was fitted using least squares regression similar to Eq.~\ref{eq:costfunction}, yielding the cost function as
\begin{equation}
\mathcal{L}(a) = \sum_{\Phi_i \in S_\Psi} (X^{\mathrm{exact}}_{\Phi_i} - f_{\mathrm{NCE}}(X^{\mathrm{noisy}}_{\Phi_i},k_{\Phi_i},a))^2.
\end{equation}
By inputting $X_{\Psi}^{\mathrm{noisy}}$ and setting $k = n$, the model predicts the mitigated value $X_{\Psi}^{\mathrm{predicted}}$,
\begin{equation}
X_\Psi^{\mathrm{predicted}} = f_{\mathrm{NCE}}(X_\Psi^{\mathrm{noisy}},n,a).
\end{equation}

\subsubsection{Resource scaling and mitigation overhead}
For a target circuit on $Q$ qubits and depth $D$, CDR-type mitigation introduces overhead through (i) the number of additional training circuits evaluated on hardware, (ii) the classical cost of generating near-Clifford labels for those circuits, and (iii) the total shot cost for estimating the relevant observables. Let $N$ denote the number of training circuits actually run on hardware (in addition to the target circuit). 

The quantum workload scales linearly in $N$ in the sense that the total number of circuit executions is $(N+1)$ times that of the unmitigated experiment. If the target observable is an energy $E=\sum_{j=1}^{L} c_j \langle P_j\rangle$ with $L$ measured Pauli terms (or commuting groups), and $S_j$ shots are used to estimate each $\langle P_j\rangle$, then a simple upper bound on the total sampling cost is $S_{\mathrm{tot}} \;=\; (N+1)\sum_{j=1}^{L} S_j$ for traditional and ES CDR.
The multi-$k$ variant NCE-CDR replaces $(N+1)$ by $\bigl(1+\sum_{k\in\mathcal{K}} N_s(k)\bigr)$ when separate training sets of sizes $N_s(k)$ are collected at several refinement values $k\in\mathcal{K}$. 

The classical burden is dominated by near-Clifford simulation of the (training) circuits: writing $C_{\mathrm{sim}}(Q,D,k)$ for the cost of classically estimating the ideal value of a depth-$D$ circuit with refinement parameter $k$, near-Clifford methods typically scale polynomially in the Clifford part (hence polynomially in $Q$ and $D$ for fixed structure) but \emph{exponentially} in the non-Clifford ``budget'' controlled by $k$. Hence, we model the classic cost as
$C_{\mathrm{sim}}(Q,D,k) \;=\; \mathrm{poly}(Q,D)\,\exp\!\bigl(\alpha k\bigr)$,
for some implementation-dependent constant $\alpha>0$. Consequently, the classical labeling cost is $\mathcal{O}\!\left(N\,C_{\mathrm{sim}}(Q,D,k)\right)$ for standard CDR, $\mathcal{O}\!\left(M\,C_{\mathrm{sim}}(Q,D,k)\right)$ for ES-CDR when simulating a pool of $M$ candidates and selecting $N<M$ to run on hardware, and $\mathcal{O}\!\left(\sum_{k\in\mathcal{K}} N_s(k)\,C_{\mathrm{sim}}(Q,D,k)\right)$ for multi-$k$ training. In the worst case, mitigation can become exponentially expensive if maintaining accuracy requires $k$ to grow with $Q$ and/or $D$; in practice, the regime of interest is when useful mitigation is achieved with modest $k$ (so that $C_{\mathrm{sim}}$ remains tractable) and with $N$ (and, for ES-CDR, $M$) at most mildly growing with problem size.  As a practical reference for the choice of $k$, state-of-the-art classical simulation algorithms have been shown to efficiently simulate circuits with on the order of $50$ qubits and around $60$ non-Clifford gates without requiring high-performance computing resources~\cite{PRXQuantum.3.020361,Bravyi2019simulationofquantum}.

In summary, our two extensions inherit the same qualitative scaling properties as in the original CDR framework\cite{Czarnik2021errormitigation}: the dominant classical overhead is still driven by the near-Clifford refinement parameter $k$ (entering exponentially through $C_{\mathrm{sim}}(Q,D,k)$), and the dominant quantum overhead by the number of training circuits $N$ (entering linearly through the total number of circuit evaluations and hence the total shot budget). Accordingly, in what follows we assess our extensions by comparing mitigation performance at fixed resource drivers, i.e., at the same $k$ when focusing on classical cost and at the same $N$ when focusing on quantum cost, and ask whether they deliver improved mitigation under these matched-cost conditions.

\section{Computational Details}
\label{sec:comp}

For all calculations, including VQE and tUPS ansatz preparation, the quantum computational software package \texttt{SlowQuant}~\cite{slowquant} was used.  \texttt{PySCF}~\cite{sun2017pythonbasedsimulationschemistryframework} was used for the Hartree-Fock reference state and \texttt{Qiskit}~\cite{qiskit} was used as the backend for quantum emulation with a noise model.
The Jordan–Wigner mapping~\cite{JordanWignerMapper} is adopted throughout the work. The Ansatz was constructed using the tiled Unitary Product State (tUPS) Ansatz~\cite{burton2024accurategateefficientquantumansatze} and for the classical optimization, the gradient-free \texttt{Rotosolve} algorithm~\cite{Ostaszewski_2021} was first applied, followed by further optimization using the gradient-based optimizer \texttt{SLSQP}~\cite{kraft1988slsqp} as provided by the \texttt{SciPy} library~\cite{scipy}.

The test system considered is the H$_4$ molecule in a rectangular geometry, with horizontal and vertical bond lengths of $1.5$ \r{A} and $1.8$ \r{A}, respectively. In the minimal basis set STO-3G~\cite{10.1063/1.1672392} chosen here, this corresponds to a full space of $4$ electrons in $4$ spatial orbitals (i.e.,\ 8 qubits). Two different tUPS Ans{\"a}tze with $2$ and $3$ layers were employed. All CDR implementations are performed for the energy evaluation on the optimized circuit parameters obtained from VQE in ideal settings.  Details of the noise-free and noisy ground state energies, as well as the corresponding absolute errors obtained under this configuration, are provided in Table~\ref{tab:energy}. The number of 2-qubit gates before and after transpilation, as well as the number of non-Clifford parameters, for our 2-layer and 3-layer circuits are shown in Table~\ref{tab:gates}. 

\begin{table}[H]
  \centering
  \caption{Noise-free and noisy ground state energies, along with their corresponding absolute energy errors, for H$_4(4,4)$ using 2- and 3-layer tUPS circuits, respectively.}
  \label{tab:energy}
  \begin{tabular}{|c|c|c|c|}
    \hline
     & noise-free energy [Ha] & noisy energy [Ha] & absolute energy error [Ha] \\ \hline
    2-layer & $-$3.712209 & $-$3.103227 & 0.608982 \\ \hline
    3-layer & $-$3.712497 & $-$2.927799 & 0.784699 \\ \hline
  \end{tabular}
\end{table}

\begin{table}[htbp]
  \centering
  \caption{Number of 2-qubit gates before and after transpilation, as well as number of non-Clifford parameters $n$ in our 2- and 3-layer tUPS circuits, respectively.}
  \label{tab:gates}
  \begin{tabular}{|c|c|c|c|}
    \hline
     & 
     2-qubit gates & 2-qubit gates (transp.) & $n$ \\ \hline
    2-layer & 120 & 270 & 18 \\ \hline
    3-layer & 180 & 405 & 27 \\ \hline
  \end{tabular}
\end{table}

All simulations were performed in the infinite shot limit to focus the study on device noise. To simulate noisy environments, a noise model from the 133-qubit fake backend \textit{FakeTorino}, provided by the \texttt{Qiskit} \textit{Fake Provider}~\cite{fakeprovider}, was applied to the primitive backend \textit{AerSimulator}~\cite{aersimulator}. This backend emulates essential hardware characteristics of a real quantum device, including the coupling map, basis gates, and noisy qubit properties such as decoherence times and error rates. As the test sets are created by randomly drawing circuits out of the full permutation space $C(n,k)$, each mitigation procedure was repeated $20$ times to compute the mean absolute error and the standard deviation. For the CDR implementations, all regressions were performed using an analytical multiple linear regression implemented with the \texttt{LinearRegression} class from \texttt{scikit-learn}~\cite{scikit-learn}.
\section{Results}
\label{sec:results}

Next, we present our results of, firstly, a comprehensive benchmark of the meta parameters of CDR using the H$_4$ molecular system with the tUPS Ansatz, and, second, showcase our enhancement strategies.

\subsection{Effect of training set parameters on traditional CDR}
\label{subsec:CDRRe}

To understand the effect of the various test set parameters in CDR, we start by presenting in detail the effect of (i) sampling size $N$, and (ii) the amount of non-Clifford parameters $k$. This is done for the H$_4$ test system using (iii) both linear and quadratic models and (iv) with and without biasing. 

\subsubsection{Sampling size}
\par
\begin{figure}[H]
  \centering
  \includegraphics[width=1\textwidth, clip, trim=0 565 0 0]{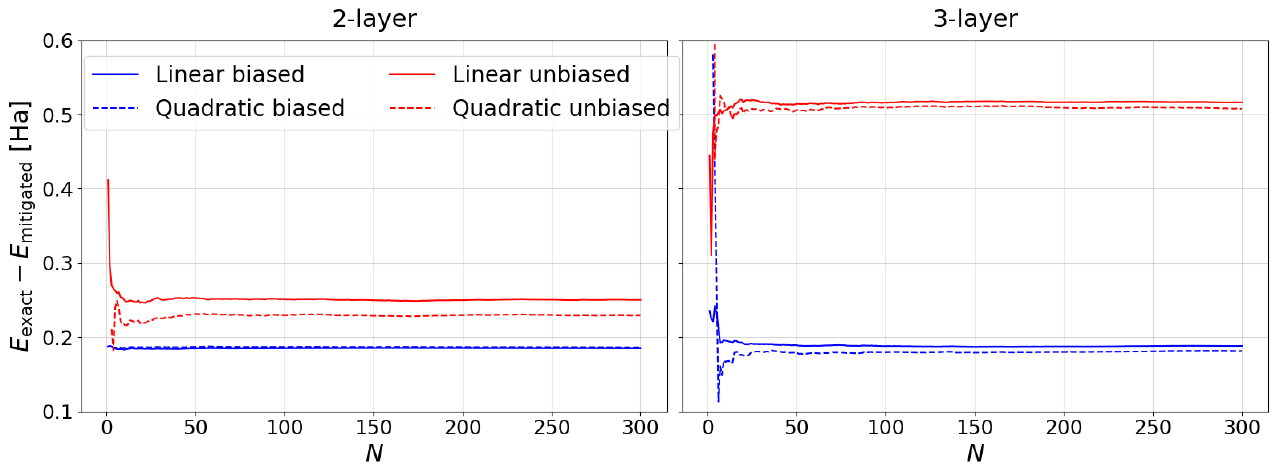}
\caption{Energy differences of noise-free and CDR-mitigated ground state energies for H$_4(4,4)$ with a tUPS Ansatz of layers $L = 2$ and $3$. The dependence of the mitigation with respect to the training set size $N$ is shown. For each system a linear (solid lines) and quadratic (dashed lines) model was used as well as with (blue lines) and without (red lines) biasing. The number of non-Clifford parameters was set to $k=4$ and $k=6$ for $L=2$ and $L=3$, respectively.}
  \label{fig:ZBNscan}
\end{figure}
\par
As a first step, we study the effect of sample size $N$, in conjunction with biasing (on and off) and linear and quadratic regression models, on the mitigated ground state energy results for a $L=2$ and $L=3$ layer tUPS Ansatz. The simulations were carried out with $N$ ranging from $1$ to $300$, and $k$ was fixed at $4$ for the 2-layer tUPS and $6$ for the 3-layer case. The results are shown in Fig.~\ref{fig:ZBNscan} in terms of the absolute energy errors as functions of the number of regression samples $N$. 
Across all four setups — namely (i) no biasing with linear model (solid red line), (ii) no biasing with quadratic model (dashed red line), (iii) biasing with linear model (solid blue line), and (iv) biasing with quadratic model (dashed blue line) — the absolute error is seen to plateau rapidly once approximately $N\!\approx\!50$ regression samples have been supplied; beyond this point only slight improvements are gained by further enlarging the training set.
In terms of the individual parameters under investigation, biasing has the largest impact; clearly, outperforming unbiased (zeroing) approach for both layers and regression types (with the average absolute errors, in the stable regime, reduced by approximately $0.05\,$Ha for the 2-layer ansatz and by as much as $0.3\,$Ha for the 3-layer Ansatz when biasing is applied). The results obtained with linear and quadratic models are very similar, indicating that the choice between them has little impact on the mitigation performance.
The corresponding standard deviations ($\sigma$) are shown in Fig.~\ref{supp-fig:Nscan_std} and show a similar trend regarding the four setups with linear and quadratic biased being converged in the std for $N\!\approx\!50$ with values of $\sigma <0.005\,$Ha and $\sigma <0.01\,$Ha for $2$ and $3$ layers, respectively. 
However, in our system, the absolute errors remain relatively large (above $0.1$ Ha) even for $N>50$, indicating limitations in the predictive accuracy of the current regression models. 

\subsubsection{Non-Clifford gates}

\begin{figure}[H]
  \centering
  \includegraphics[width=1\textwidth, clip, trim=0 565 0 0]{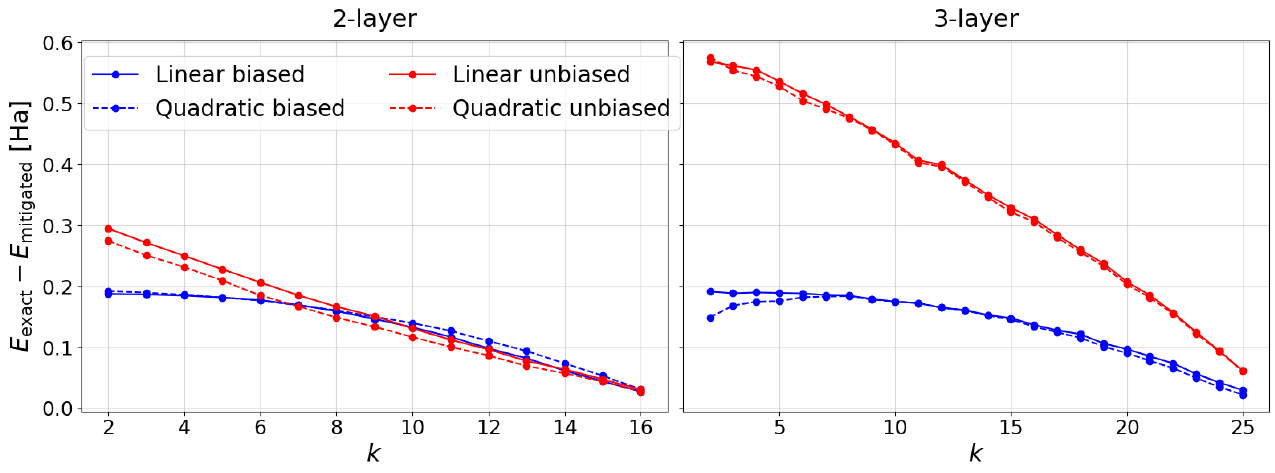}
  \caption{Energy differences between noise-free and mitigated ground state energies for H$_4$ plotted as a function of the number of non-Clifford parameters $k$, using both 2-layer and 3-layer tUPS 
  Ans{\"a}tze. For each system, both linear (solid lines) and quadratic (dashed lines) regression models were performed, with and without biasing (blue and red lines, respectively). CDR was performed using $N = 153$ samples in the training datasets for both 2- and 3-layer tUPS ansatze.}
  \label{fig:ZBkscan}
\end{figure}

In the previous section, the number of non-Clifford parameters, $k$, was set to a small number ($4$ or $6$). As this still shows large absolute errors, the effect of increasing the number of non-Clifford parameters will now be investigated. Here, $k$ is ranging from $2$ to $16$ for 2-layers and from $2$ to $25$ for 3-layers, while the number of samples is fixed at $N= 153$ ($C(18,2)=153$, the maximum number of different near-Clifford circuits with $k=2$ for 2-layer Ansatz). 
Fig.~\ref{fig:ZBkscan} illustrates a decreasing trend in energy difference for both the 2-layer and 3-layer tUPS cases as $k$ increases. This behavior is expected, since a larger number of non-Clifford parameters retained in the circuit brings it closer to the original circuit, meaning that the sample states $\ket{\Phi_i}$ in the set of training states $S_{\Psi}$ gradually approach the target state $\ket{\Psi}$. Although increasing $k$ improves the accuracy of the mitigation results, it also exponentially increases the computational cost of running the circuits on a classical simulator~\cite{PhysRevA.70.052328}, thus necessitating a trade-off.

From the perspective of the $k$-scan (Fig.~\ref{fig:ZBkscan}), the choice between linear (solid lines) and quadratic (dashed lines) models continues to make little difference in the mitigation results, similar to the observation from the $N$-scan (Fig.~\ref{fig:ZBNscan}). 
However, the difference in mitigation performance between biased (blue lines)and zero circuit (red lines) preparations is strongly dependent on $k$ and the circuit depth. For the 2-layer tUPS case, zero preparation exhibits a more linear behavior, whereas biased preparation resembles a non-linear trend. At $k = 1$, the absolute error of the zero-mitigated result is $1.5$ times larger than that of the biased-mitigated result, while this gap gradually narrows and eventually becomes negligible at $k = 6$ and beyond. For the 3-layer case, the biased result outperforms for all values of $k$. At $k = 1$, the absolute error of the zero-mitigated result is three times larger than that of the biased-mitigated result and this gap gradually decreases as $k$ increases. Unlike the 2-layer case, it does not nearly disappear until $k$ approaches the maximum number of non-Clifford parameters. These results further highlight the advantage of biased circuit preparation. However, even at $k = 16$ or $25$, where the circuits are close to optimal, our mitigated results still struggle to approach milliHartree precision.

\subsection{Energy sampling}
After having understood the impact of various parameters in the test set for traditional CDR, we next apply our enhancement idea of energy sampling, outlined in section~\ref{ES}. Here, all circuits are prepared using the biased preparation strategy, and the regression models remain both linear and quadratic.

\begin{figure}[H]
  \centering
  \includegraphics[width=1\textwidth, clip, trim=0 565 0 0]{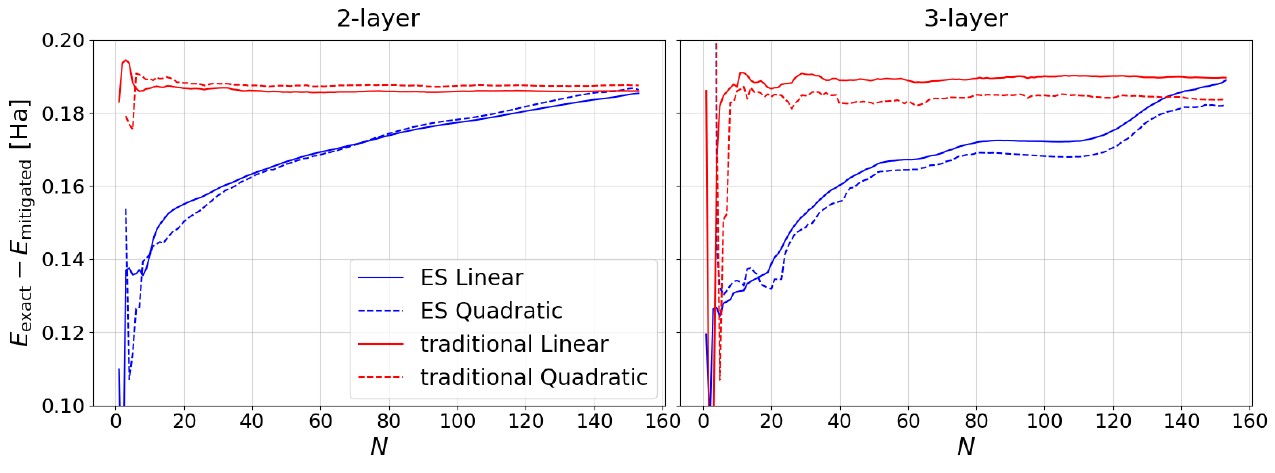}
  \caption{Absolute energy errors of the CDR-mitigated results with both 2-layer and 3-layer tUPS Ans{\"a}tze, plotted as a function of training set size $N$. CDR was implemented with (ES, blue lines) and without (traditional, red lines) our energy sampling strategy. All training circuits were generated using biased preparation, with both linear (solid lines) and quadratic (dashed lines) models employed. For the ES-CDR, $N$ lowest-energy circuits were selected from a biased data pool of $M = 153$ samples. For $L=2$ and $L=3$, the number of the non-Clifford parameters $k$ is fixed at $4$ and $6$, respectively.}
  \label{fig:MoN}
\end{figure}

In Fig.~\ref{fig:MoN}, the influence of selecting low-energy samples on the accuracy of CDR is demonstrated by drawing an increasingly large subset of the $N$ lowest-energy circuits from the biased training pool of $M=153$ circuits with $k=4$ for 2-layers and $k=6$ for 3-layers. The traditional CDR baseline is trained on just $N$ circuits, i.e., the same number of training samples but chosen without the low-energy filter. As the energy sampling occurs via classic simulation of the near-Clifford circuits, the training set and thus the quantum workload for traditional and ES-CDR are identical.

The absolute error shows that when $N$ is small, the regression results with energy sampling (ES) significantly outperform traditional CDR, in some cases halving the absolute errors. As more lowest-energy circuits are included in the regression dataset, the absolute error gradually increases, eventually converging with traditional CDR results when all circuits are included (i.e., $N = 153$ out of $M = 153$). The sharp fluctuations for small $N$ values indicate instabilities in that range. This is further supported by the standard deviations shown in Fig.~\ref{supp-fig:M_out_of_153_std}. The standard deviations decrease rapidly with increasing $N$ and eventually stabilize at around $N=40$. 

\begin{figure}[H]
  \centering
  \includegraphics[width=1\textwidth, clip, trim=0 565 0 0]{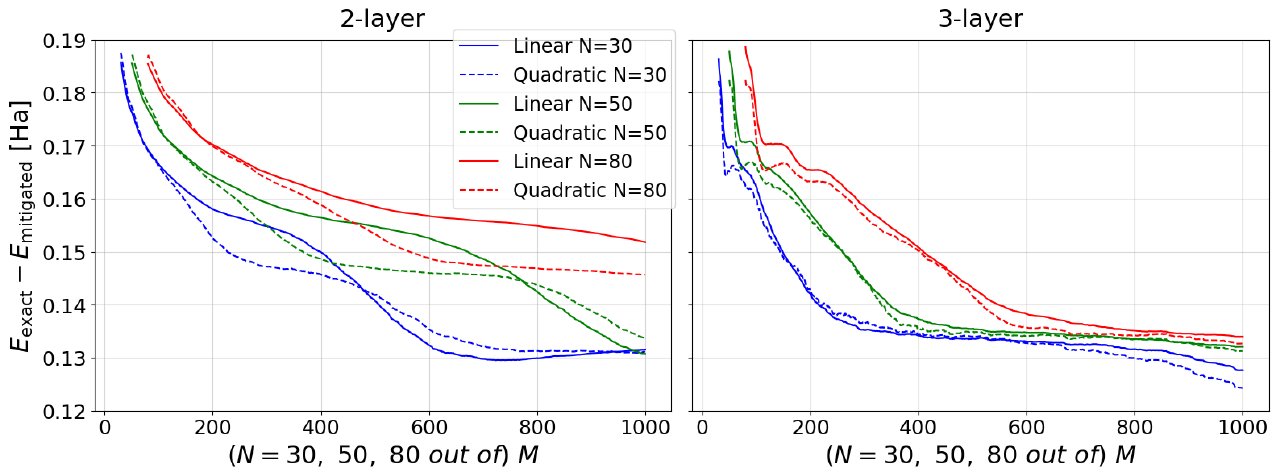}
  \caption{Absolute energy errors for a fixed number of selected samples $N$, with the data pool size $M$ varying from $N$ to $1000$. For both 2-layer and 3-layer tUPS, $N$ is fixed at 30 (blue), 50 (green), and 80 (red). For 2-layer and 3-layer cases, the number of the non-Clifford parameters $k$ are fixed at $4$ and $6$, respectively.
  }
  \label{fig:FMoN}
\end{figure}

Finally, we examined the impact of varying the biased training data pool size $M$ on the mitigation performance while keeping $N$ (the number of low energy samples for the test set) fixed. In Fig.~\ref{fig:FMoN}, the number of selected lowest-energy circuits was chosen as $N = 30$ (blue), $50$ (green), and $80$ (red) and the pool size $M$ was varied between $M=[N,1000]$. Having $M=N$ equates to performing traditional CDR. Crucially, we see that for each fixed value of $N$, the absolute error decreases with increasing the pool size $M$. This is because selecting the same number of lowest-energy samples from a larger training pool tends to result in samples with overall lower energies, thereby guiding the regression model to produce lower energy predictions. Therefore, the energy sampling (ES CDR) consistently improves on traditional CDR.

\subsection{Non-Clifford Extrapolation}
\begin{figure}[H]
  \centering
  \includegraphics[width=1\textwidth, clip, trim=0 565 0 0]{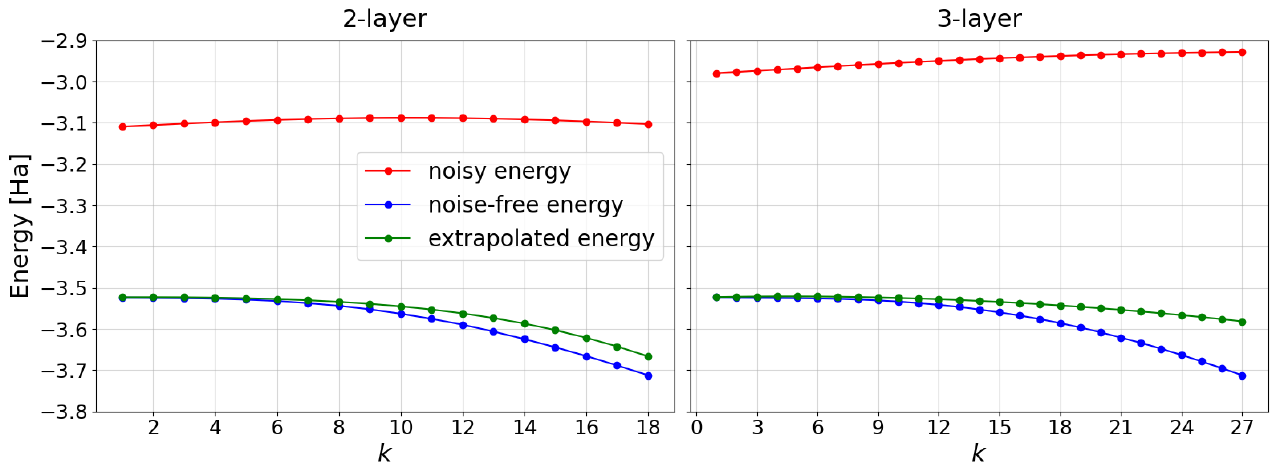}
  \caption{Mean noisy (red lines) and noise-free (blue lines) ground state energies as well as extrapolated results (green lines) for H$_4(4,4)$ obtained from a noisy quantum simulator, ideal quantum simulation and our extrapolation model, respectively. For the 2-layer case, data with $k$ ranging from $1$ to $4$ is used to train the model, while for the 3-layer case, $k$ ranges from $1$ to $6$. For each $k$ value, 1000 samples were used for statistics.
  }
  \label{fig:kenergy}
\end{figure}

To motivate our proposal of non-Clifford extrapolation (see section~\ref{NCE}), a detailed study of the relationship between the noisy and noise-free ground state energies in dependence of $k$ is presented in Fig.~\ref{fig:kenergy}.

Here, we show how both the noise-free (blue line) and noisy values (red line) evolve with increasing $k$ for the 2- and 3-layer tUPS An{s\"a}tze. To obtain better statistics, for each value of $k$, $1000$ circuits were generated. Both the 2-layer and 3-layer cases exhibit the same trend: the noise-free energy shows a clear decreasing with increasing $k$. This behavior is expected, as increasing $k$ brings the sampled circuits closer to the exact circuit with optimal parameters, thereby yielding lower energy values. In contrast, the noisy energy changes only slightly with $k$, which may be due to the fact that although the circuits approach the optimal ones, the presence of noise distorts their execution, preventing the measurement outcomes from reflecting a clear dependence on $k$.

In traditional and ES CDR, only samples corresponding to a single $k$ value are used as the training set. That is, the regression model was trained on data from a small $k$, and then the noisy value at $k = n$ was input into the model to predict the corresponding noise-free value. However, as shown above, the noisy and noise-free energies exhibit different trends as $k$ increases, indicating that their relationship varies with $k$. As a result, a model trained solely on data from a single small $k$ value cannot accurately predict the noise-free value at $k = n$. To address this issue, the Non-Clifford extrapolation CDR (NCE CDR) approach is proposed. It performs regression using samples collected at multiple small $k$ values, allowing the model to learn how the relationship between noisy and noise-free values evolves with $k$, and thereby extrapolate this relationship to $k = n$ for more accurate prediction of the target energy. This method requires the $k$ value of each sample to be included as an additional input feature in the regression ansatz, as defined in Equation \ref{eq:kansatz}.

\begin{figure}[H]
  \centering
  \includegraphics[width=1\textwidth, clip, trim=0 565 0 0]{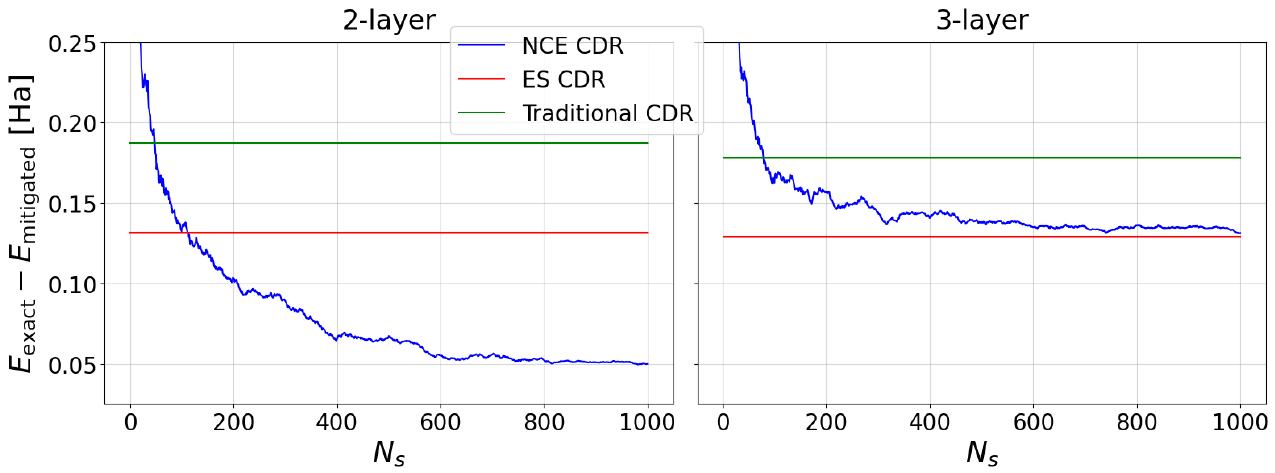}
  \caption{Absolute energy errors of the CDR mitigated results with Non-Clifford extrapolation (NCE, blue lines), plotted as a function of $N_s$ (sampling uniformly $N$ times per $k$ value). Samples with $k$ ranging from $1$ to $4$ for the 2-layer, and from $1$ to $6$ for the 3-layer tUPS are used as the training set. For comparison, the absolute errors of the converged and stable results obtained by the traditional CDR (green lines) at $N = 40$ as well as ES CDR (red lines) at $N = 40$ and $M = 1000$ are also presented. The $k$ values of traditional and ES CDR are fixed at $4$ and $6$ for 2- and 3-layer tUPS, respectively.}
  \label{fig:kfeature}
\end{figure}

Consistent with previous analyses, we begin by examining how the number of samples $N$ influences the mitigation performance. Crucially, as we sample now at various $k$ values, we sample uniformly $N$ times per $k$ value, unless the maximum number of samples that can be generated for a given $k$ is smaller than $N$. In such cases, $N_s$ is set to the maximum value, given by $C(n, k)$, i.e.\ $N_s=\mathrm{min}(N,C(n,k))$. Fig.~\ref{fig:kfeature} presents the absolute energy error plotted against the number of training samples per $k$ value, $N_s$, where the training set comprises circuits with $k$ from $1$ to $4$ for the 2-layer tUPS and from $1$ to $6$ for the 3-layer tUPS. A similar trend can be observed for both the 2-layer and 3-layer results: when only a small number of samples per $k$ are used, the absolute error decreases rapidly. As $N_s$ increases, the error gradually levels off and enters a plateau. In the case of the 2-layer circuit, the absolute error decreases and converges to approximately $0.05$ Ha, entering a relatively stable regime as $N_s$ increases further. For the 3-layer circuit, the error converges to approximately $0.13$ Ha when $N_s$ approaches $600$. Compared with traditional and ES CDR, our NCE CDR method outperforms both approaches in the 2-layer case, reducing the absolute error by about $0.13$ Ha and $0.08$ Ha, respectively. In the 3-layer case, it still shows an improvement over traditional CDR, although its performance does not consistently surpass that of ES CDR, with slightly larger errors observed.

The difference in performance can be understood by visualizing the effect of extrapolation or rather how good the extrapolation scheme is able to recover the true energy at $k=n$, see green line in Fig.~\ref{fig:kenergy}. For the 2-layer case (trained on $k=[1,4]$), the extrapolated results closely match the noise-free values until around $k=10$ with slightly larger deviations beginning to appear beyond this point and a final deviation at $k=n=18$ of $0.05\,$Ha. For the 3-layer case (trained on $k=[1,6]$), the differences between the extrapolated and noise-free values begin to emerge and increase gradually after $k=11$, reaching approximately $0.13\,$Ha at $k=27$ (optimal circuit). This indicates that, for the 3-layer circuit, training the model using only data with $k=[1,6]$ does not provide sufficient information about the relationship between the noisy and noise-free values. To address this, samples with larger $k$ values can be incorporated into the training set (see below). In Fig.~\ref{supp-fig:2layer_nce} and Fig.~\ref{supp-fig:3layer_nce}, we show examples with varying the training set and see that they exhibit a similar trend to the results above, and extending the upper limit of $k$ to $8$ leads to only minor improvements in performance. 

While NCE CDR demonstrates advantages over other CDR variants in certain cases, it comes at the cost of larger test sets because circuits for a $k$ range instead of a single $k$ are included and the convergence with respect to $N$ is slower in NCE CDR. For example, at a $N_s$ of $40$ (giving converged results in traditional and ES CDR), we observe larger errors for NCE CDR and standard deviations (see SI) of $0.19$ Ha and $0.23$ Ha for 2- and 3-layer cases, respectively. Additionally, the cumulative size of the test set is $240$. 
However, it is important to realize that while NCE CDR exhibit larger costs, it can systematically improve the energy estimates, suggesting the potential for achieving higher accuracy.

\begin{figure}[H]
  \centering
  \includegraphics[width=1\textwidth, clip, trim=0 565 0 0]{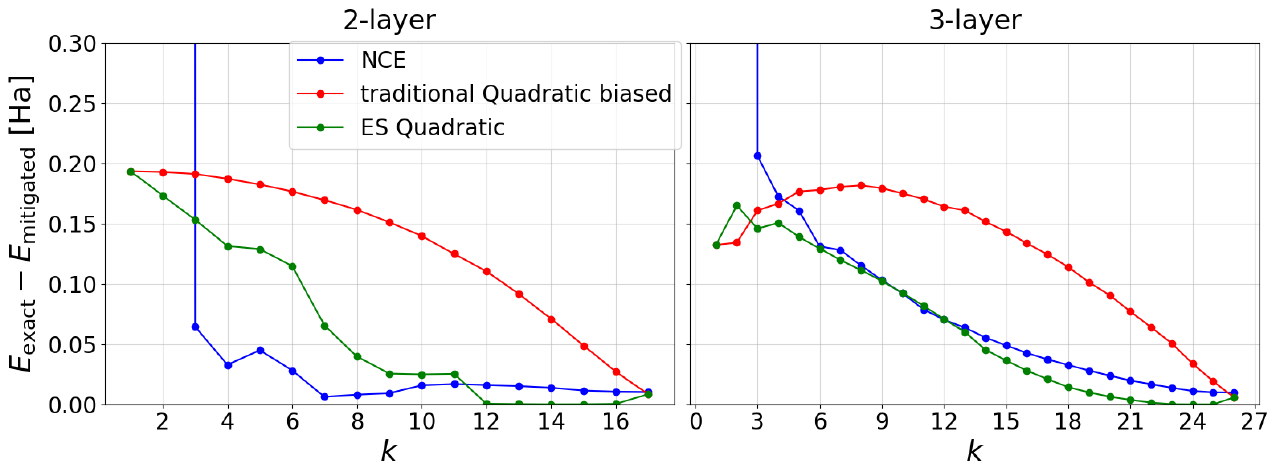}
  \caption{Absolute energy errors of the CDR mitigated results with NCE CDR (blue lines), plotted as a function of $k$. For both the 2- and the 3-layer tUPS, the number of non-Clifford parameters in the training circuits ranges from $1$ to $k$, while $N_s$ is fixed at $1000$. For comparison, the absolute errors of the mitigated results obtained with traditional (red lines) and ES CDR (green lines) are also plotted against $k$. In traditional CDR, $N$ is fixed at $1000$, whereas in ES CDR, $N=40$ and $M=1000$ are used.}
  \label{fig:kscan}
\end{figure}

After comparing the performance of different CDR variants in terms of training set size, we now analyze how the absolute energy errors behave as the number of non-Clifford parameters $k$ increases across all methods. Fig.~\ref{fig:kscan} shows the absolute energy errors of CDR mitigated results obtained using NCE CDR, traditional CDR, and ES CDR as functions of $k$. In both the 2- and 3-layer tUPS cases, the NCE CDR (blue lines) exhibits a more rapid reduction in error with increasing $k$, achieving lower overall absolute errors compared to traditional CDR (red lines). For the 2-layer system, NCE CDR reaches nearly zero error beyond $k \approx 6$, significantly outperforming both baselines. In the 3-layer case, although the advantage is less pronounced, NCE CDR still maintains competitive accuracy—comparable to ES CDR (green lines). These results demonstrate that including training circuits with multiple $k$-values enables NCE CDR to achieve more accurate extrapolations, highlighting its potential as a systematic improvement over traditional CDR approach.
\section{Conclusion}
\label{sec:con}

In this work, we systematically investigated  and improved upon the Clifford Data Regression (CDR) error mitigation strategy for quantum chemistry simulations on 
NISQ devices. As benchmark we used the H$_4$ molecule in the STO-3G minimal basis with the tiled unitary product state (tUPS) Ansatz of varying depth and applied CDR to mitigate noise-induced errors in the ground state energies estimation. To this end, the hyperparameters of the traditional CDR method were thoroughly studied, and two enhancements, namely Energy sampling (ES) and Non-Clifford extrapolation (NCE), were proposed to improve its accuracy.

Our numerical study on traditional CDR showed that biasing training circuits toward the target state consistently lowers absolute errors while  quadratic regression models can sometimes yield slight improvements over linear models, but the effect is generally small and inconsistent. Convergence studies showed that with respect to the training set size, we observed rapid convergence after a few tens of circuits and the expected systematic improvement with increase in number of non-Clifford parameters.

Energy sampling CDR (ES CDR) increases the classically simulated pool whilst keeping the quantum simulation and training set size consistent with traditional CDR by selecting only the lowest energy solution out of the larger classical pool. We show that this biasing systematically improves the error mitigation capabilities.

Non-Clifford extrapolation CDR (NCE CDR) extends the regression model by including an increasing number of non-Clifford parameters as an additional input, enabling the model to learn how the noisy–ideal mapping evolves as the circuit approaches the exact non-Clifford parameter solution. Our numerical study shows that NCE CDR outperform traditional CDR. Its advantage over ES CDR is inconsistent and comes at the cost of requiring more training samples. 

In summary, both enhancement strategies outperform their traditional counterpart across a scan of circuit depth and number of non-Clifford parameters. ES CDR is cost-effective best option as it can even outperform NCE CDR at very low quantum costs, while NCE CDR opens the possibility of further improved results at higher costs.

Further studies will explore more advanced regression models for NCE CDR and investigate a combination of ES and NCE CDR. 
\section{Acknowledgments}
\label{sec:acknowle}

We acknowledge the support of the Novo Nordisk Foundation (NNF) for the focused research project “Hybrid Quantum Chemistry on Hybrid Quantum Computers” (Grant No. NNFSA220080996).
\newpage
\bibliography{references}
\end{document}


\renewcommand{\thefigure}{S\arabic{figure}}
\renewcommand{\thesection}{S\arabic{section}}
\renewcommand{\thepage}{S\arabic{page}}
\maketitle
\pagebreak
\section{Additional Figures}
\label{SI}

\begin{figure}[H]
  \centering
  \includegraphics[width=1\textwidth, clip, trim=0 565 0 0]{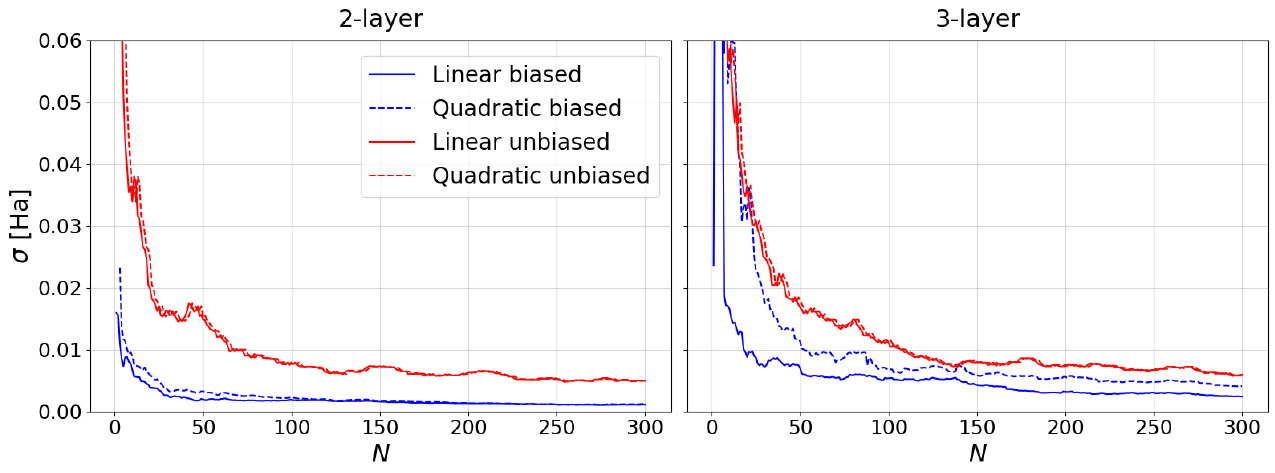}
\caption{Standard deviation ($\sigma$) of CDR-mitigated ground state energies for H$_4(4,4)$ with a tUPS Ansatz of layers $L = 2$ and $3$. The dependence of the mitigation with respect to the training set size $N$ is shown. For each system a linear (solid lines) and quadratic (dashed lines) model was used as well as with (blue lines) and without (red lines) biasing. The number of non-Clifford parameters was set to $k=4$ and $k=6$ for $L=2$ and $L=3$, respectively.}
  \label{fig:Nscan_std}
\end{figure}

\begin{figure}[H]
  \centering
  \includegraphics[width=1\textwidth, clip, trim=0 565 0 0]{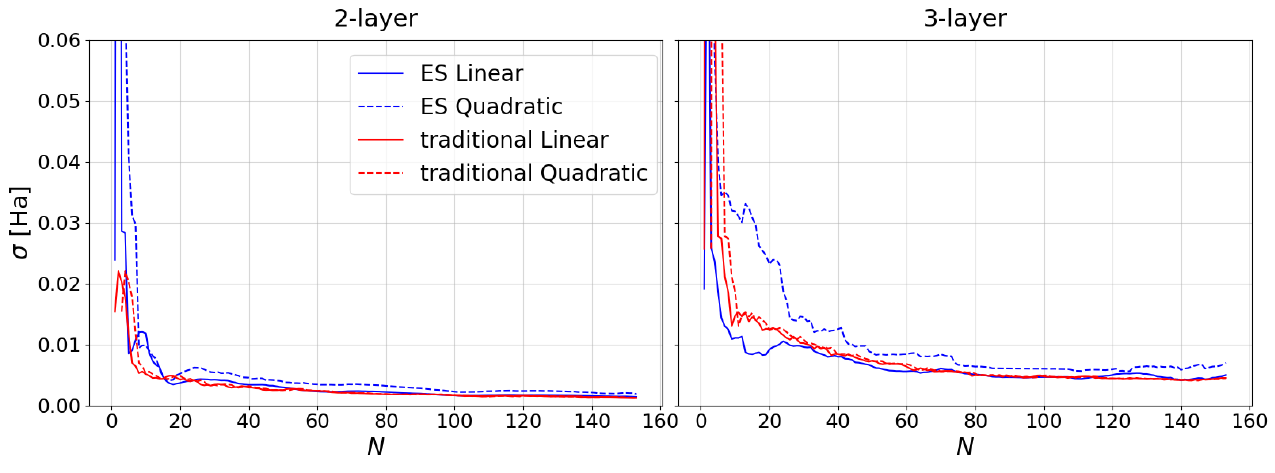}
  \caption{Standard deviation ($\sigma$) of the CDR-mitigated results with both 2-layer and 3-layer tUPS ansatze, plotted as a function of training set size $N$. CDR was implemented with (ES, blue lines) and without (traditional, red lines) our energy sampling strategy. All training circuits were generated using biased preparation, with both linear (solid lines) and quadratic (dashed lines) models employed. For the ES-CDR, $N$ lowest-energy circuits was selected from a biased data pool of $M = 153$ samples. For $L=2$ and $L=3$, the number of the non-Clifford parameters $k$ are fixed at $4$ and $6$, respectively.}
  \label{fig:M_out_of_153_std}
\end{figure}

\begin{figure}[H]
  \centering
  \includegraphics[width=1\textwidth, clip, trim=0 565 0 0]{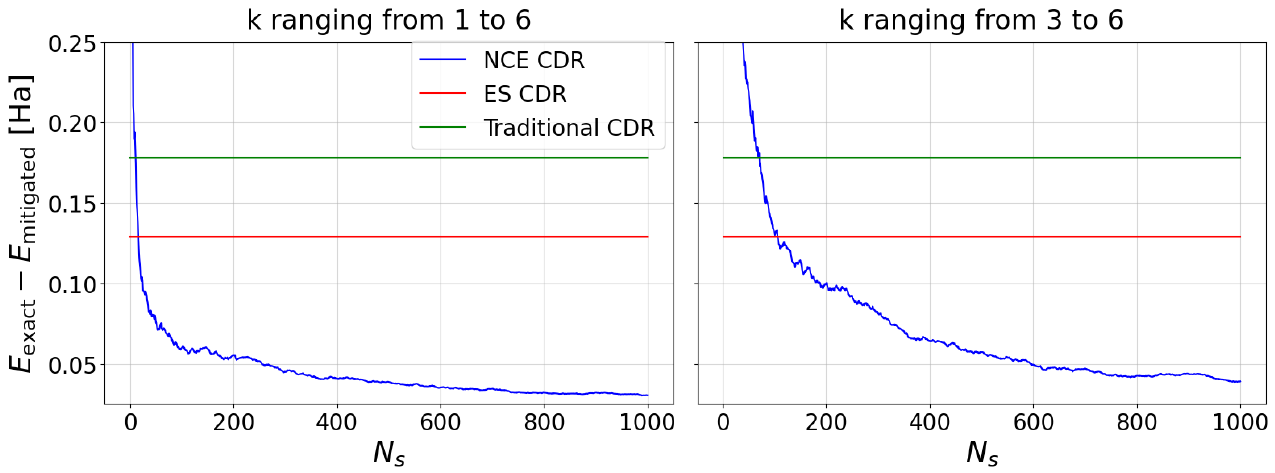}
  \caption{Absolute energy errors of the CDR mitigated results with Non-Clifford extrapolation (NCE, blue lines), plotted as a function of $N_s$ (sampling uniformly $N$ times per $k$ value). Samples with $k$ ranging from $1$ to $6$ and $3$ to $6$ for the 2-layer tUPS are used as the training set. For comparison, the absolute errors of the converged and stable results obtained by the traditional CDR (green lines) at $N = 40$ as well as ES CDR (red lines) at $N = 40$ and $M = 1000$ are also presented. The $k$ values of traditional and ES CDR are fixed at $6$.}
  \label{fig:2layer_nce}
\end{figure}

\begin{figure}[H]
  \centering
  \includegraphics[width=1\textwidth, clip, trim=0 640 0 0]{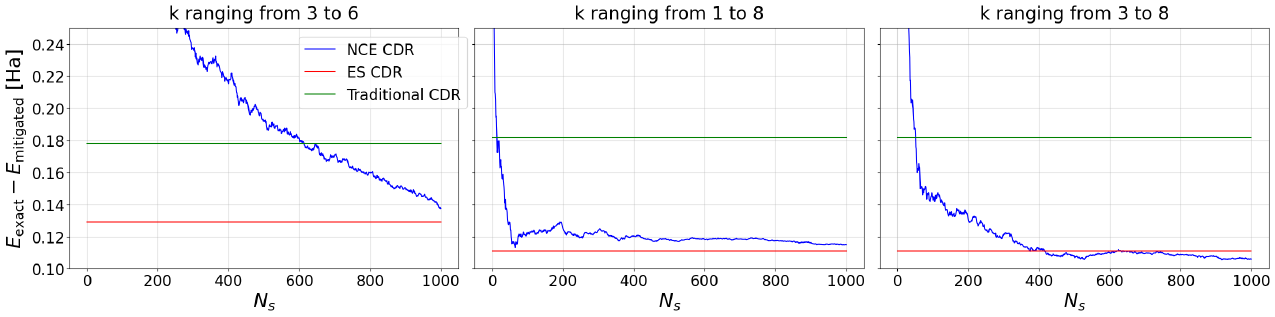}
  \caption{Absolute energy errors of the CDR mitigated results with Non-Clifford extrapolation (NCE, blue lines), plotted as a function of $N_s$ (sampling uniformly $N$ times per $k$ value). Samples with $k$ ranging from $3$ to $6$, $1$ to $8$ and $3$ to $8$ for the 3-layer tUPS are used as the training set. For comparison, the absolute errors of the converged and stable results obtained by the traditional CDR (green lines) at $N = 40$ as well as ES CDR (red lines) at $N = 40$ and $M = 1000$ are also presented. The $k$ values of traditional and ES CDR are fixed at $6$ for $k$ ranging from $3$ to $6$ and $8$ for $k$ ranging from $1$ to $8$ and $3$ to $8$, respectively.}
  \label{fig:3layer_nce}
\end{figure}